\documentstyle[12pt]{article}
\newcommand{\beq}{\begin{equation}}
\newcommand{\eeq}{\end{equation}}
\newcommand{\bea}{\begin{eqnarray}}
\newcommand{\eea}{\end{eqnarray}}

\newcommand{\gsim}{\lower.7ex\hbox{$
\;\stackrel{\textstyle>}{\sim}\;$}}
\newcommand{\lsim}{\lower.7ex\hbox{$
\;\stackrel{\textstyle<}{\sim}\;$}}

\newcommand{\eod}{\end{document}}

\def\ot{{\bf T}}
\def\cp{{\bf CP}}
\def\cpt{{\bf CPT}}

\newcommand{\NP}{New Physics}

\begin{document}
\thispagestyle{empty}
\vspace*{-22mm}

\begin{flushright}
UND-HEP-07-BIG\hspace*{.08em}08\\
%hep-ph/0703132\\
\today

\end{flushright}
\vspace*{1.3mm}

\begin{center}
{\LARGE{\bf
On the Brink of Major Discoveries in Weak Charm Decays --\\ a Bismarckian Chance to Make History
}}
\footnote{Invited lecture given at MENU07, Sept. 10 - 14, 2007, FZ J\"ulich, Germany}
\vspace*{19mm}

{\Large{\bf I.I.~Bigi}} \\
\vspace{7mm}

{\sl Department of Physics, University of Notre Dame du Lac}
\vspace*{-.8mm}\\
{\sl Notre Dame, IN 46556, USA}\\
{\sl email: ibigi@nd.edu}

\vspace*{10mm}

{\bf Abstract}\vspace*{-1.5mm}\\
\end{center}

\noindent
The recently obtained strong evidence for $D^0 - \bar D^0$ oscillations from the $B$ factories provides an important benchmark in our search for New Physics in charm transitions. While the theoretical verdict on the observed values of $x_D$ and $y_D$ is ambiguous -- they could be fully generated by SM dynamics, yet could contain also a sizable contribution from New Physics -- such oscillations provide a new stage for \cp~violation driven by New Physics. After emphasizing the unique role of charm among up-type quarks, I describe in some detail the \cp~phenomenology for charm decays.

%\tableofcontents

\vspace{0.5cm}

\noindent 
{\bf Prologue}

\vspace{0.3cm}

While the sub-division of research in fundamental physics is the natural result of history, it makes eminent sense to consider periodically, whether its specific form is still optimal. This conference has assembled a majority of people from Middle Energy Physics with a sizable contingent from High Energy Physics. I have noticed during the talks that both communities share a dual goal in their research, namely to validate theoretical control over the Standard Model (SM) dynamics -- most talks at this conference are devoted to this topic -- and to search for New Physics. The latter is conducted at three frontiers: (i) The `High Energy Frontier', which will be pushed into new territories with the operation of the LHC beginning next year; (ii) the `High Accuracy Frontier', for which the analysis of the muon's 
$g-2$ is the most impressive example, and (iii) the `High Sensitivity Frontier' best illustrated by the probe of \cp~invariance in the decays of $K$, $B$ and $D$ mesons. Items (ii) and (iii) represent indirect searches for \NP. The greatly enhanced experimental 
sensitivities for $K$, $B$ and $D$ transitions and the typically moderate size contributions anticipated from \NP~mandate that the aspects of high accuracy and high sensitivity get combined in those studies, although not on the level of $g-2$. It is on this new combined frontier where in my view the Middle Energy and High Energy communities can and should form new alliances. 

%%%%%%%%%%%%%%%%%%
\section{Introduction}
%%%%%%%%%%

While the study of strange dynamics was instrumental in the formation of the SM and that of charm transitions central for the SM being accepted, the analysis of $B$ decays almost completed its validation through the establishment of CKM dynamics as the dominant  
source of the observed \cp~violation; `almost', since the Higgs boson has not been observed yet. Now the race is on to see which of these areas together with top quark decays -- will first reveal an incompleteness of the SM in flavour dynamics. If the evidence for 
$D^0-\bar D^0$ oscillations with $x_D$, $y_D \sim 0.005 - 0.01$ listed below gets confirmed, then the detailed probe of \cp~symmetry in charm decays is a close second behind the race leader, namely the even more detailed study of  $B$ decays. 

The signal for $D^0 - \bar D^0$ oscillations is a {\em tactical} draw: while the observed values $x_D$ and $y_D$ might be generated by SM forces alone, they could contain large contributions from \NP. Yet a {\em strategic} breakthrough is in sight: future probes of 
\cp~symmetry in $D$ decays can decide, whether \NP~manifests itself in charm transitions. I would like to draw a historical analogy based on my personal experience. Sanda and myself had been talking about large 
\cp~asymmetries in certain $B$ decays \cite{BS80} 
without much resonance -- till $B_d - \bar B_d$ oscillations were found by the ARGUS collaboration in 1987 \cite{ARGUS87}, 
i.e. twenty years ago. 
Since the oscillation parameter $x_D$ is about two orders of magnitude smaller than $x_B$, \cp~asymmetries in $D$ decays will be much smaller than what was found in $B$ decays. However we should note that the `background' from SM dynamics is even tinier -- meaning the \NP~signal to `SM noise' ratio might actually be considerably better than in $B$ decays. I would also count on our experimentalists having become more experienced and thus being able to extract smaller signals. 

%%%%%%%%%%%%
\section{\NP~Scenarios \& the Uniqueness of Charm}
%%%%%%%%%%%

\NP~in general induces flavour changing neutral currents (FCNC). It was actually one of the formative challenges for the SM to 
reproduce the observed great suppression of strangeness changing neutral currents. One should note that the couplings of FCNC could be substantially less suppressed for up-type than for down-type quarks. This actually happens in some models which `brush the dirt' of FCNC in the down-type sector under the `rug' of the up-type sector. 
{\em Among up-type quarks it is only charm that allows the full range of probes for 
FCNC and New Physics in general}: (i) Top quarks decay {\em before} they can hadronize \cite{RAPALLO}. Without top {\em hadrons} 
$T^0 - \bar T^0$ oscillations cannot occur. This limits our options 
to search for \cp~asymmetries, since one cannot call on oscillations to provide the required second amplitude. 
(i) Hadrons built with $u$ and $\bar u$ quarks like $\pi^0$ and $\eta$ are their own antiparticle; thus there can be no $\pi^0 - \pi^0$ etc. oscillations as a matter of principle. Furthermore they possess so few 
decay channels that \cpt~invariance basically rules out  \cp~asymmetries in their decays.  

I will show below that only very recently have experiments reached a range of sensitivity, where one can realistically expect 
\cp~violation to show up in charm transitions. My basic contention is as follows: {\em Charm transitions are a unique portal for obtaining novel access to flavour dynamics with the experimental situation being a priori favourable apart from the absence of Cabibbo suppression.} 

%%%%%%%%%%%%%%%%
\section{On the Evidence for $D^0 - \bar D^0$ Oscillations}
%%%%%%%%%%%%%%

Oscillations are described by the normalized mass and width 
splittings:  $x_D \equiv \frac{\Delta M_D}{\Gamma_D}$, $y_D \equiv \frac{\Delta \Gamma_D}{2\Gamma_D}$. 
While the SM predicts similar numbers for $x_D$ and $y_D$ with the data showing the same trend, we should note that $\Delta M_D$ and $\Delta \Gamma_D$ reflect rather different dynamics:  
$\Delta M_D$ is produced by {\em off}-shell transitions making it naturally sensitive to \NP~ 
unlike $\Delta \Gamma_D$, which is generated by {\em on}-shell modes.  

%%%%%%%%%%
\subsection{The Data}
%%%%%%%%%%

I will list here only those data that show an effect with the strongest significance.  
\begin{itemize}
\item 
Finding different lifetimes in the decays of neutral $D$ mesons constitutes an unequivocal 
manifestation of $D^0$ oscillations. 
BELLE obtains a 3.2 $\sigma$ signal for a difference in the effective lifetimes for $D^0 \to K^+K^-$ and $D^0 \to K^-\pi^+$  \cite{BELLEOSC1}: 
\beq 
y_{CP} = \frac{\tau (D^0 \to K^-\pi^+)}{\tau (D^0 \to K^+K^-)} - 1 = (1.31 \pm 0.32 \pm 0.25) \cdot 10^{-2}   
\label{BELLEYCP}
\eeq
In the limit of \cp~invariance (a good approximation for charm decays as explained later)  
the two mass eigenstates of the $D^0 - \bar D^0$ complex are \cp~eigenstates as well. 
$D^0 \to K^+K^-$ yields the width for the \cp~even state and $D^0 \to K^-\pi^+$ the one averaged 
over the \cp~even and odd states and thus:  $y_{CP} = y_D = \frac{\Delta \Gamma_D}{2\bar \Gamma _D}$. 
\item 
The selection rule $\Delta C = \Delta S$ is violated in the SM by doubly Cabibbo suppressed 
$c \to d \bar s u$ transitions (DCSD). By analyzing the decay rate evolution as a function of (proper) time, one can disentangle the two sources for `wrong-sign' kaons: 
\beq
\frac{{\rm rate}(D^0(t) \to K^+\pi^-)}{{\rm rate}(D^0(t)\to K^-\pi^+)} 
 =  \frac{|T(D^0\to K^+\pi^-)|^2}{|T(D^0\to K^-\pi^+)|^2}  \cdot   
[1 +Y_{K\pi}(t\Gamma_D) +  
Z_{K\pi}(t\Gamma_D)^2)]  
\eeq 
\beq
Y_{K\pi} \equiv   
\frac{y_D}{{\rm tg}^2\theta_C }  
{\rm Re}\left(\frac{q}{p}\hat \rho_{K\pi}  
\right)  + \frac{x_D}{{\rm tg}^2\theta_C}  
{\rm Im}\left(\frac{q}{p}\hat \rho_{K\pi}  
\right)    , \, 
 Z_{K\pi}  \equiv    
\frac{x_D^2 +y_D^2}{4{\rm tg}^4\theta_C} |\hat \rho _{K\pi}|^2
\eeq  
where we have used the notation 
\beq
\frac{T(\bar D^0 \to K^+\pi^-)}{T(D^0 \to K^+\pi^-)} = \frac{1}{{\rm tg}^2\theta_C}\hat \rho _{K\pi}
\eeq
to emphasize that the {\em non}-oscillation amplitude is doubly Cabibbo suppressed. 
The first and third term in the square brackets represent the pure DCSD and  oscillation terms, respectively, 
and the second one their interference.  The latter receives a nonzero contribution from  
Im$\left(\frac{p}{q}\frac{\hat \rho_{K\pi}}{|\hat \rho_{K\pi}|}\right)$,  
if there is a {\em weak} phase, which leads to CP violation as discussed  
later, and/or if a {\em strong} phase is present due to different FSI  
in $D^0\to K^+\pi^-$ and $\bar D^0\to K^+\pi^-$. One has to allow for  
such a difference since the latter is a pure $\Delta I=1$ transition, 
while the former is given by a combination of an enhanced  
$\Delta I=0$ and a suppressed $\Delta I=1$ amplitude.  This strong phase 
$\delta$ can be absorbed into modified expressions for $x_D$ and $y_D$: 
\beq 
x_D^{\prime} \equiv x_D {\rm cos}\delta + y_D {\rm sin}\delta \; , \; 
y_D^{\prime} \equiv - x_D {\rm sin}\delta + y_D {\rm cos}\delta 
\label{XYPRIME}
\eeq 
yielding $\left( x_D^{\prime}\right) ^ 2 + \left( y_D^{\prime}\right) ^ 2 = 
x_D^2 + y_D^2$ to obtain
\beq  
Y_{K\pi}  =   
\frac{y_D^{\prime}}{{\rm tg}^2\theta_C }  \left| \frac{q}{p}\hat \rho_{K\pi}  \right| 
\eeq
Since a priori there is no reason why $\delta$ should be particularly small, one better keeps 
the difference between $(x_D,y_D)$ and $(x_D^{\prime},y_D^{\prime})$ in mind. BABAR has found \cite{BABAROSC}
\beq 
y_D^{\prime} = (0.97 \pm 0.44 \pm 0.31)\cdot 10^{-2} \; , \; 
(x_D^{\prime})^2 = (-2.2 \pm 3.0 \pm  2.1)\cdot 10^{-4} 
\eeq
representing a 3.9 $\sigma$ signal for $[y_D^{\prime}, (x_D^{\prime})^2] \neq [0,0]$ due to the 
correlations between $y_D^{\prime}$ and $ (x_D^{\prime})^2$. 
This reaction is a prime candidate for revealing \cp~violation due to new Physics, and we will discuss 
it in more detail later. 
\item 
Analyzing the time {\em dependent} Dalitz plot for $D^0(t) \to K_S\pi^+\pi^-$ BELLE finds 
\cite{BELLEOSC2}
\beq 
x_D \equiv \frac{\Delta M_D}{\Gamma_D} = (0.80 \pm 0.29 \pm 0.17)\cdot 10^{-2} \; , \; 
y_D = (0.33 \pm 0.24 \pm 0.15) \cdot 10^{-2} \; , 
\eeq
which amounts to a 2.4 $\sigma$ signal for $x_D \neq 0$. 
\footnote{BELLE extracts from its analysis the ratio between doubly Cabibbo suppressed and favoured 
amplitudes for kaon resonances of increasing mass. The trend of 
strongly increasing ratios given in their analysis can, however, hardly be correct on theoretical 
grounds \cite{BIGILUIS}. The values extracted for $x_D$ and $y_D$ are probably not very sensitive to this shortcoming 
in their Dalitz plot model.}

\end{itemize}
While all these findings are most intriguing, they do not (yet) establish the existence of $D^0$ oscillations. 
A `preliminary' average by the Heavy Flavour Averaging Group over all relevant data yields 
 5 $\sigma$ significance for $[x_D,y_D] \neq [0,0]$ with $x_D$ and $y_D$ in the range 
0.5 - 1\% -- and the caveat that averaging over 
the existing data sets has to be taken with quite a grain of salt at present due to the 
complicated likelihood functions. 

Establishing $D^0 - \bar D^0$ oscillations would provide a novel insight into 
flavour dynamics. After having discovered oscillations in {\em all three} mesons built from {\em down}-type quarks -- $K^0$, $B_d$ and $B_s$ -- it would be the first observation of oscillations with 
{\em up}-type quarks; it would also remain the only one (at least for three-family scenarios), as explained above.

%%%%%%%%%%%%
\subsection{The Inconclusive Theoretical Interpretation}
%%%%%%%%%%

The history of the predictions on $D^0$ oscillations does not provide a tale of 
consistently sound judgment by theorists, when they predicted 
$x_D \leq {\rm few}\times 10^{-4}$. Yet 
scientific progress is not made by majority vote, although that codifies it in the end. 
Within the SM two reasons combine to make $x_D$ and $y_D$  
small in contrast to the situation for $B^0 - \bar B^0$ and  
$K^0 - \bar K^0$ oscillations:  
(i) The amplitude for $D^0 \leftrightarrow \bar D^0$ transitions is twice  
Cabibbo suppressed and therefore  
$x_D$, $y_D$ $\propto {\rm sin}^2 \theta _C$. The amplitudes for  
$K^0 \leftrightarrow \bar K^0$ and $B^0 \leftrightarrow \bar B^0$  
are also twice Cabibbo and KM suppressed -- yet  
so are their decay widths.  
(ii) Due to the GIM mechanism\index{GIM mechanism} 
one has $\Delta M = 0 = \Delta \Gamma$ in  
the limit of flavour symmetry. Yet $K^0 \leftrightarrow \bar K^0$ is  
driven by $SU(4)_{Fl}$ breaking characterised by $m_c^2 \neq m_u^2$,  
which represents no suppression on the usual hadronic scales. In  
contrast $D^0 \leftrightarrow \bar D^0$ is controled by  
$SU(3)_{fl}$ breaking. Having two Cabibbo suppressed classes of decays one concludes 
for the overall oscillation strength: 
$\frac{\Delta M_D}{\bar \Gamma_D}, \; \Delta \Gamma_D \sim  
\; SU(3)_{fl}\; {\rm breaking}  
\times 2 {\rm sin}^2\theta_C   < {\rm few} \times 0.01$.  
The proper description of $SU(3)_{fl}$ {\it breaking} thus becomes the  
central issue. While $x_D \stackrel{<} \sim y_D$ is a natural finding in the SM, $x_D \ll y_D$ would not be although it cannot be ruled out. For if $D^0 \to f \to \bar D^0$ can occur for an on-shell final state $f$ thus contributing to $\Delta \Gamma_D$, then 
$D^0 \to "f" \to \bar D^0$ is possible for $"f"$ taken off-shell; i.e., $\Delta M_D$ and $\Delta \Gamma_D$ are related by a dispersion 
relation. 

One can invoke two complementary treatments to evaluate $\Delta M_D$ and $\Delta \Gamma_D$ 
in the SM. One  approach \cite{BUDOSC} 
relies on an operator product expansion (OPE) in terms of quark and gluon operators including 
nonperturbative contributions, which yield contributions in powers of $m_s/m_c$ and $\mu_{\rm had}/m_c$, where 
$m_s$ and $m_c$ denote the mass of strange and charm quarks, respectively, and $\mu_{\rm had}$ hadronic condensates. 
Terms of order $m_s^2\mu_{\rm had}^4/m_c^6$ yield the largest contributions rather than the formally leading term $m_s^4/m_c^4$, and one finds 
\beq 
x_D(SM)|_{OPE}, \; y_D(SM)|_{OPE} \sim {\cal O}(10^{-3}) 
\eeq 
with a preference for $|x_D(SM)|_{OPE} < y_D(SM)|_{OPE}$. It is unlikely that this prediction can be sharpened numerically. It should also be noted that limitations to quark-hadron duality due to the proximity of hadronic thresholds could enhance in particular $y_D$.  

The authors of 
Refs.\cite{FALK1} find similar numbers, albeit in a quite different approach:  (i) They estimate 
the amount of   
$SU(3)_{fl}$ breaking for $\Delta \Gamma_D$ from phase space differences alone for two-, three- and 
four-body $D$ modes and arrive at  $y_D(SM) \sim 0.01$. The proximity of hadronic thresholds is reflected in this number; it thus attempts to incorporate limitations in quark-hadron duality in the 
language of the OPE treatment. (ii) They infer $x_D$ from $y_D$ via a 
dispersion relation arriving at $0.001 \leq |x_D(SM)| \leq 0.01$ with $x_D$ and $y_D$ of opposite sign.  

%Concerning the predictions one has to distinguish carefully 
%between two similar sounding questions: 
%(i) "What are the {\em most likely} values for $x_D$ and $y_D$ within the SM?" 
%The answer as given above: For both $\sim {\cal O}(10^{-3})$. 
%(ii) "How large could $x_D$ and $y_D$ {\em conceivably} be within the SM?" 
%The answer: One cannot rule out $10^{-2}$. 

A priori it would have been conceivable to measure $y_D \ll x_D \sim {\rm few}\times 0.01$ thus establishing an indirect manifestation of \NP. This has not happened: we are in a grey zone, where the observed strengths of both $y_D$ and $x_D$ might be produced by SM forces alone -- or could contain significant contributions from \NP. Even in the former case one should probe 
these oscillations as accurately as possible first establishing $[x_D,y_D] \neq [0,0]$ and then determining $x_D$ vs. $y_D$. 

A future theoretical breakthrough might allow us to predict $\Delta M_D|_{SM}$ and $\Delta \Gamma_D|_{SM}$ 
more accurately and thus resolve the ambiguity in our interpretation, but I would not count on it. Rather than wait for that to happen the community should become active in the catholic tradition of `active repentance' and search for \cp~violation in $D$ decays.

%%%%%%%%%%%%
\section{\cp~Violation -- the Decisive Stage}
%%%%%%%%%%%%%%

Probing \cp~invariance for manifestations of \NP~is not a `wild goose chase'. For we know that CKM dynamics is completely irrelevant for baryogenesis; i.e., we need \cp~violating \NP~to understand the Universe's observed baryon as a dynamically generated quantity rather than an arbitrary initial value. Charm decays offer several pragmatic advantages in such searches: 
(i) While we do not know how to reliably compute the strong phase shifts 
required for direct \cp~violation to emerge in partial widths, we can expect them to be in general large, 
since charm decays proceed in a resonance domain. (ii) The branching ratios into relevant modes 
are relatively large. (iii) \cp~asymmetries can be linear in New Physics amplitudes thus enhancing sensitivity to the latter. (iv)  The `background' from known physics is small: within the SM the effective weak phase is highly diluted, namely 
$\sim {\cal O}(\lambda ^4)$. {\em Without} oscillations only 
direct \cp~violation can occur, and it can 
arise only in singly Cabibbo suppressed transitions, where one  
expects them to reach no better than the 0.1 \% level; significantly larger values would signal New Physics.  
{\em Almost any} asymmetry in Cabibbo 
allowed or doubly suppressed channels requires the intervention of New Physics, since -- in the absence of oscillations -- there is only one weak amplitude. The exception are channels containing 
a $K_S$ (or $K_L$) in the final state like $D \to K_S \pi$. There are two sources for a 
\cp~asymmetry from known dynamics: (i) Two transition amplitudes are actually involved, namely 
a Cabibbo favoured and a doubly suppressed one, $D \to \bar K^0\pi$ and $D \to K^0\pi$, respectively. 
Their relative weak CKM phase is given by $\eta A^2 \lambda ^6 \sim {\rm few} \cdot 10^{-5}$, which seems to be well beyond observability. (ii) While one has $|T(D \to \bar K^0 \pi)| = 
|T(\bar D \to K^0 \pi)|$, the well-known \cp~impurity $|p_K|\neq |q_K|$ in the $K_S$ wave function introduces a difference 
between $D^{0,+}\to K_S\pi^{0,+}$ and $\bar D^{0,-}\bar K_S \pi^{0,-}$ of 
$\frac{|q_K|^2 - |p_K|^2}{|q_K|^2 + |p_K|^2} = (3.32 \pm 0.06)\cdot 10^{-3}$ \cite{CICERONE}. 

With oscillations on an observable level -- and it seems $x_D$, $y_D$ $\sim 0.005 - 0.01$ satisfy 
this requirement -- the possibilities for \cp~asymmetries proliferate. Even if \NP~is not the main engine for  
$\Delta M_D$, it could well be the leading source of \cp~violation in ${\cal L}(\Delta C=2)$. 
This would be analogous to the very topical case of $B_s$ oscillations. 
$\Delta M (B_s)$ has been observed to be 
consistent with the SM prediction within mainly theoretical uncertainties; yet since those are 
still sizable, we cannot rule out that New Physics impacts $B_s - \bar B_s$ oscillations 
significantly. This issue, which is unlikely to be resolved theoretically, can be decided experimentally 
by searching for a time dependent \cp~violation in $B_s(t) \to \psi \phi$. For within the SM one predicts \cite{BS80} a very small asymmetry not exceeding 4\% in this transition since on the leading CKM level quarks of only the second and third family contribute. Yet in general one can expect New Physics contributions to $B_s - \bar B_s$ oscillations to exhibit a weak phase that is not particularly 
suppressed. Even if New Physics affects $\Delta M(B_s)$ only moderately, it could greatly enhance 
the time dependent \cp~asymmetry in $B_s(t) \to \psi \phi$. 
This analogy is of course qualitative rather than quantitative with 
$D^0 - \bar D^0$ oscillations being (at best) quite slow. 

%%%%%%%%%%%%%%%%
\subsection{Oscillations -- the New Portal to \cp~Violation} 
%%%%%%%%%%%%%%%

In the presence of $D^0 - \bar D^0$ oscillations 
{\em time-dependent} \cp~asymmetries 
can arise in $D^0$ decays on the Cabibbo allowed ($D^0 \to K_S\phi$, $K_S\rho^0$, $K_S\pi ^0$), 
once forbidden ($D^0 \to K^+K^-$) and doubly forbidden $(D^0 \to K^+\pi^-$) levels. Let me list just two prominent 
examples from the last two categories. Since $y_D$, $x_D \ll 1$, it suffices to give the 
decay rate evolution to first order in those quantities only (the general expressions can be found in 
Ref.\cite{CICERONE}). 
\bea 
\nonumber
\Gamma (D^0(t) \to K^+K^-) &\propto & e^{-\Gamma_1t}|T(D^0 \to K^+K^-)|^2 \times  \\
\nonumber 
&& 
\left[ 1 +y_D\frac{t}{\tau_D} \left( 1 - {\rm Re}\frac{q}{p}\bar \rho_{K^+K^-}\right) - 
x_D\frac{t}{\tau_D}{\rm Im}\frac{q}{p}\bar \rho_{K^+K^-}\right] \\
\nonumber 
\Gamma (\bar D^0(t) \to K^+K^-) &\propto & e^{-\Gamma_1t}|T(\bar D^0 \to K^+K^-)|^2\times 
\\ 
&& 
\left[ 1 +y_D\frac{t}{\tau_D} \left( 1 - {\rm Re}\frac{p}{q} \frac{1}{\rho_{K^+K^-}}\right) - 
x_D\frac{t}{\tau_D}{\rm Im}\frac{p}{q}\frac{1}{\rho_{K^+K^-}} \right]
\label{DKK} 
\eea
The usual three types of \cp~violation can arise, namely the direct and indirect types -- 
$|\bar \rho_{K^+K^-}| \neq 0$ and $|q|\neq |p|$, respectively -- as well as the one involving 
the interference between the oscillation and direct decay amplitudes -- 
Im$\frac{q}{p}\bar \rho_{K^+K^-}\neq 0$ leading also to Re$\frac{q}{p}\bar \rho_{K^+K^-}\neq 1$. 
Assuming for simplicity $|T(D^0 \to K^+K^-)| = |T(\bar D^0 \to K^+K^-)|$ 
\footnote{CKM dynamics is expected 
to induce an asymmetry not exceeding 0.1\%.} and $|q/p| = 1- \epsilon_D$ one has 
$(q/p)\bar \rho_{K^+K^-} = (1-\epsilon_D) e^{i\phi_{K\bar K}}$ and thus 
\beq 
A_{\Gamma} = \frac{\Gamma (\bar D^0(t) \to K^+K^-) - \Gamma (D^0(t) \to K^+K^-)}
{\Gamma (\bar D^0(t) \to K^+K^-) + \Gamma (D^0(t) \to K^+K^-)} 
\simeq x_D\frac{t}{\tau_D} {\rm sin}\phi_{K\bar K} -  
y_D\frac{t}{\tau_D}\epsilon_D {\rm cos}\phi_{K\bar K}\; .  
\eeq
where I have assumed $|\epsilon_D| \ll 1$. 
BELLE has found \cite{BELLEOSC1}
\beq 
A_{\Gamma} = (0.01 \pm 0.30 \pm 0.15) \%
\eeq
While there is no evidence for \cp~violation in the transition, one should also note that 
the asymmetry is bounded by $x_D$. For $x_D$, $y_D \leq 0.01$, as indicated by the data, 
$A_{\Gamma}$ could hardly exceed the 1\% range. I.e., there is no real 
bound on $\phi_D$ or $\epsilon_D$ yet. The good news is that if $x_D$ 
and/or $y_D$ indeed fall into the 0.5 - 1 \% range, 
then any improvement in the experimental sensitivity for a \cp~asymmetry in 
$D^0(t) \to K^+K^-$ constrains New Physics scenarios -- or could reveal them 
\cite{GKN}. 

Another promising channel for probing \cp~symmetry is $D^0(t) \to K^+\pi^-$: since it is 
doubly Cabibbo suppressed, it should a priori exhibit a higher sensitivity to a 
New Physics amplitude.  Furthermore it cannot exhibit direct \cp~violation in the SM. 
With 
\beq
\frac{q}{p} \frac{T(D^0 \to K^+\pi^-)}{T(D^0 \to K^-\pi^+)}
\left[   \frac{p}{q} \frac{T(\bar D^0 \to K^-\pi^+)}{T(\bar D^0 \to K^+\pi^-)}  \right] \equiv  
 - \frac{1}{{\rm tg}^2\theta_C} (1-[+] \epsilon_D) |\hat \rho _{K\pi}|e^{-i(\delta - [+]\phi_{K\pi})}
\eeq
%\bea 
%\nonumber 
%\frac{q}{p} \frac{T(D^0 \to K^+\pi^-)}{T(D^0 \to K^-\pi^+)} &\equiv & 
% - \frac{1}{{\rm tg}^2\theta_C} (1-\epsilon_D) |\hat \rho _{K\pi}|e^{-i(\delta - \phi_{K\pi})} \\
% \frac{q}{p} \frac{T(\bar D^0 \to K^-\pi^+)}{T(\bar D^0 \to K^+\pi^-)} &\equiv & 
% - \frac{1}{{\rm tg}^2\theta_C} \frac{1}{1-\epsilon_D} |\hat \rho _{K\pi}|e^{-i(\delta + \phi_{K\pi})}
%\eea
one expresses an asymmetry as follows: 
$$  
\frac{\Gamma (\bar D^0(t) \to K^-\pi^+) - \Gamma (D^0(t) \to K^+\pi^-)}
{\Gamma (\bar D^0(t) \to K^-\pi^+) + \Gamma (D^0(t) \to K^+\pi^-)} \simeq 
$$
\beq
 \left(\frac{t}{\tau_D}\right) \left| \hat \rho_{K\pi}\right|
 \left( \frac{y_D^{\prime}{\rm cos}\phi_{K\pi}\epsilon_D - 
x_D^{\prime}{\rm sin}\phi_{K\pi}}{{\rm tg}\theta_C^2}\right) + 
 \left(\frac{t}{\tau_D}\right)^2 \left| \hat \rho_{K\pi}\right|^2 \frac{\epsilon_D(x_D^2 + y_D^2)}
 {2{\rm tg}\theta_C^4}
 \eeq
 where I have again assumed for simplicity $|\epsilon _D| \ll 1$ and {\em no direct} 
 \cp~violation. 
 
BABAR has also searched for a time dependent \cp~asymmetry in $D^0 \to K^+\pi^-$ vs. 
$\bar D^0(t) \to K^- \pi^+$, yet so far has not found any evidence for it \cite{BABAROSC}. Again, with $x_D^{\prime}$ and $y_D^{\prime}$ capped by 
about 1\%, no nontrivial bound can be placed on the weak phase $\phi_{K\pi}$. On the other hand any further increase in experimental sensitivity could reveal a signal. 

%%%%%%%%%%%%%%%%%
\subsection{\cp~Asymmetries in Final State Distributions}
%%%%%%%%%%%%% 

Decays to final states of {\em more than} two pseudoscalar or one pseudoscalar and one vector meson contain 
more dynamical information than given by their  widths; their distributions as described by Dalitz plots 
or \ot{\em -odd} moments can exhibit \cp~asymmetries that can be considerably larger than those for the 
width. All \cp~asymmetries observed so far in $K_L$ and $B_d$ decays 
except one concern partial widths, i.e. 
$\Gamma (P \to f)$ $\neq$ $\Gamma (\bar P \to \bar f)$. The one 
notable exception can teach us important lessons for future searches both in charm and $B$ decays, namely the \ot~odd moment found in $K_L\to \pi ^+ \pi ^- e^+ e^-$. Denoting by $\phi$ the angle between the $\pi^+\pi^-$ and $e^+e^-$ planes 
one has 
\beq 
\frac{d\Gamma}{d\phi}(K_L \to \pi^+\pi^- e^+e^-) = \Gamma_1 {\rm cos}^2 \phi + 
\Gamma_2 {\rm sin}^2 \phi + \Gamma_3 {\rm cos} \phi {\rm sin}\phi 
\eeq
Comparing the $\phi$ distribution integrated over two quadrants one obtains a \ot~odd moment: 
\beq 
\langle A \rangle = \frac{\int _0^{\pi/2}d\phi \frac{d\Gamma}{d\phi} - \int _{\pi/2}^{\pi}d\phi \frac{d\Gamma}{d\phi}}
{\int _0^{\pi}d\phi \frac{d\Gamma}{d\phi}}= \frac{2\Gamma_3}{\pi (\Gamma_1 + \Gamma_2)}
\label{<A>}
\eeq
$\langle A \rangle $ is measured to be $0.137 \pm 0.015$ \cite{PDG06} in full agreement with the prediction of 
$0.143 \pm 0.013$ \cite{SEGHALKL}. 
Most remarkably this large asymmetry is generated by the tiny \cp~impurity parameter $\eta_{+-}\simeq 0.0024$; i.e., the impact 
of the latter is magnified by a factor of almost a hundred -- for the price of a tiny branching ratio of about $3\cdot 10^{-7}$! 

This trading of asymmetry against branching ratio can be attempted also in the so far 
unobserved rare charm mode 
$D_L \to K^+K^- \mu^+\mu^-$, where $D_L$ denotes the \cp~odd longer lived mass eigenstate. 
The \cp~impurity parameter $\epsilon_D$ that controls $D_L \to K^+K^-$ can get enhanced by almost two orders of magnitude in the 
\ot~odd moment defined analogous to $\langle A \rangle$ in Eq.(\ref{<A>}) \cite{AYAN}. The required $D_L$ beam can be prepared through a EPR correlation \cite{EPR} in 
$e^+e^- \to \gamma ^* \to D_S D_L$ near threshold, where the shorter lived $D_S$ is tagged through 
$D_S \to K^+K^-$, $\pi^+\pi^-$. 

The same effects can be probed also by comparing the $\phi$ distributions in 
$D^0 \to K^+K^-\mu^+\mu^-$ vs.  
$\bar D^0 \to K^+K^-\mu^+\mu^-$ or in 
$D^0 \to K^+K^-\pi^+\pi^-$ vs. $\bar D^0 \to K^+K^-\pi^+\pi^-$ \cite{PEDRINI}. The aforementioned 
huge enhancement factor, however, does not arise then.

%%%%%%%%%%%%%
\subsection{\cp~Violation in Semileptonic $D^0$ Decays}
%%%%%%%%%%%%%%%

$|q/p| \neq 1$ unambiguously reflects \cp~violation in $\Delta C=2$ dynamics. It can be probed most directly in semileptonic $D^0$ decays leading to `wrong sign' leptons: 
\beq 
a_{SL}(D^0) \equiv \frac{\Gamma (D^0(t) \to l^-X) - \Gamma (\bar D^0 \to l^+X)}
{\Gamma (D^0(t) \to l^-X) + \Gamma (\bar D^0 \to l^+X)} = 
\frac{|q|^4 - |p|^4}{|q|^4 + |p|^4} 
\eeq
The corresponding observable has been studied in semileptonic decays of neutral $K$ and $B$ mesons. With $a_{SL}$ being controlled by $(\Delta \Gamma/\Delta M){\rm sin}\phi_{weak}$, 
it is predicted to be small in both cases, albeit for different reasons: 
(i) While $(\Delta \Gamma_K/\Delta M_K) \sim 1$ one has sin$\phi_{weak}^K \ll 1$ leading to 
$a_{SL}^K = \delta _l  \simeq (3.32 \pm 0.06)\cdot 10^{-3}$ as observed. 
(ii) For $B^0$ on the other hand one has 
$(\Delta \Gamma_B/\Delta M_B)\ll1$ leading to $a_{SL}^B < 10^{-3}$. 

For $D^0$ both $\Delta M_D$ and $\Delta \Gamma_D$ are small, yet 
$\Delta \Gamma_D/\Delta M_D$ is not: present data indicate it is about unity or even larger; 
$a_{SL}$ is given by the smaller of $\Delta \Gamma_D/\Delta M_D$ or its inverse multiplied by 
sin$\phi_{weak}^D$, which might not be that small: i.e., while the rate for `wrong-sign' leptons is small in semileptonic decays of neutral 
$D$ mesons, their \cp~asymmetry might not be at all, if New Physics intervenes to generate 
$\phi_{weak}^D$.  

%%%%%%%%%%%%%%%
\section{Conclusions and Outlook}
%%%%%%%%%%%%

It is of great importance to firmly establish the existence of $D^0 - \bar D^0$ oscillations and determine $x_D$ vs. $y_D$. 
My main message is that we must go after \cp~violation in charm transitions in all of its possible manifestations, both 
time dependent and independent, in partial widths and final state distributions, and on all Cabibbo levels down to the 
$10^{-3}$ or even smaller level. The present absence of any \cp~asymmetry is not telling. Comprehensive and detailed studies of charm decays provide a novel and possibly unique window onto flavour dynamics. 

For that purpose we need the statistical muscle of LHCb. Charm studies constitute a worthy challenge to LHCb, for which 
$D^0 \to K^+K^-$, $\pi^+\pi^-$, $K^+\pi^-$, $K^+K^-\mu^+\mu^-$ represent good channels. Yet I feel we have to go after even more 
statistics and more channels. This brings me to my second main message adapted from Cato the Elder: 

\begin{center} 
"Ceterum censeo fabricam super saporis esse faciendam!" 

"Moreover I advise a super-flavour factory has to be built!" 
\end{center}

Bismarck, who exhibited a flexibility concerning morality similar to Cato's, once declared: 
" ... it is the role of the statesman to grab the mantle of history when he feels it passing by." Likewise it is the task of the physicist to make the greatest use of a special gift from Nature. $D^0 - \bar D^0$ oscillations are such a gift; it is therefore our duty to make the most complete use of it -- and there is fame within our grasp.

\vspace{0.5cm}

{\bf Acknowledgments:} This work was supported by the NSF under the grant number PHY-0355098. 
I am grateful to Profs. Krewald and Machner for inviting me to this fine meeting and for 
making it possible that most participants could stay in the great city of Aachen.

\vspace{4mm}

%%%%%%%%%%%%%%%%

\end{document}